\documentclass[]{raa}            

\usepackage{graphicx,times}             
\usepackage{amsmath}
\usepackage{arydshln}
\usepackage{multirow}
\usepackage{subfig}
\begin{document}

   \title{Interference coupling analysis based on a hybrid method: application for radio telescope system\,$^*$
\footnotetext{\small $*$ Supported by the National Basic Research Program of China and the National Natural Science Foundation of China.}
}

   \volnopage{Vol.0 (200x) No.0, 000--000}      
   \setcounter{page}{1}          

   \author{Qing-Lin Xu
      \inst{1}
   \and Yang Qiu
      \inst{1}
   \and Jin Tian
      \inst{1}
   \and Qi Liu
      \inst{2}
   }

   \institute{School of Mechano-Electronic Engineering, Xidian University, Xi'an 710071, China; {\it qlxu@stu.xidian.edu.cn}\\
        \and
             Xinjiang Astronomical Observatory, Chinese Academy of Sciences, Urumqi 830011, China\\
   }

   \date{Received~~2009 month day; accepted~~2009~~month day}

\abstract{ Working in a way that passively receives the electromagnetic radiation of celestial body, the radio telescope is easily disturbed by external radio frequency interference (RFI) as well as the electromagnetic interference (EMI) generated by electric and electronic components operating at telescope site. The quantitative analysis for these interferences must be taken into account carefully for further electromagnetic protection of the radio telescope. In this paper, based on the electromagnetic topology (EMT) theory, a hybrid method that combines Baum-Liu-Tesche (BLT) equation and transfer function is proposed. In this method, the coupling path of radio telescope is divided into strong coupling and weak coupling sub-paths, and the coupling intensity criterion is proposed by analyzing the conditions that BLT equation simplifies to transfer function. According to the coupling intensity criterion, the topological model of a typical radio telescope system is established. The proposed method is used to solve the interference response of the radio telescope system by analyzing the subsystems with different coupling modes respectively and then integrating the responses of the subsystems as the response of the entire system. The validity of the proposed method is verified numerically. The results indicate that the proposed method, compared with the direct solving method, reduces the difficulty and improves the efficiency of the interference prediction.
\keywords{telescopes --- methods: analytical --- methods:
numerical --- waves}
}

   \authorrunning{Q.-L. Xu et al. }            
   \titlerunning{Interference coupling analysis on radio telescope}  

   \maketitle

%
%
\section{Introduction}           
\label{sect:intro}

While radio astronomy service has been assigned the use of many frequency bands, the radio telescope is still faced with the radio frequency interference (RFI), such as unwanted out-of-band transmissions from adjacent and nearby bands (Waterman~\cite{wate84}). On the other hand, the electromagnetic interference (EMI) generated by devices at radio telescope site is also a major threat (Ambrosini et al.~\cite{ambr10}). When these interferences couple into the radio telescope system through multiple paths, the ability of radio astronomical observations would be reduced and even the radio telescope cannot work normally.

In order to protect the radio telescope from RFI, the threshold levels of detrimental interference in radio astronomy bands have been given in Recommendation ITU-R RA.769. Therefore, radio telescope sites are usually chosen in remote area so as to satisfy the threshold levels (Driel~\cite{driel09}; Umar et al.~\cite{umar14}). Further, shielded cabinets and Faraday cages have been used to attenuate EMI (Abeywickrema et al.~\cite{abey15}). In order to determine the level of attenuation of mitigation techniques, it is necessary to analyze these interferences quantificationally. Some measurements have been performed to estimate the absolute interference level (Bolli et al.~\cite{boll13}; Hidayat et al.~\cite{hida14}). However, it usually needs a long time to carry out the measurements for a radio telescope system because of its large volume and complex structure. Even some measurements cannot be performed because the test conditions cannot be met. Therefore, the interference prediction based on numerical and analytical methods has been investigated by different authors.

In order to predict the interference coupled into a complex system, numerical methods, such as finite-difference time-domain method (Georgakopoulos et al.~\cite{geor01}) and method of moments (Audone \& Balma~\cite{audo89}; Araneo \& Lovat~\cite{aran09}), have been used to calculate the interference response. However, they often require much computing time and memory because of the detailed mesh generation in electrically large system (Tzeremes et al.~\cite{tzer04}), moreover it is difficult to investigate the variation of interference response with design parameters. Analytical formulations, although approximate, provide a better efficiency, and enable the variation with design parameters to be investigated, but it is difficult to handle complex geometry and material (Nie et al.~\cite{nie11}). Among these methods, electromagnetic topology (EMT) owns the specific advantage to analyze electromagnetic interaction with complex electronic systems. The EMT method was first proposed by Baum~(\cite{baum74}) and later improved by Tesche~(\cite{tesc78}). In this method, interaction sequence diagram (Tesche \& Liu~\cite{tesc86}) and Baum-Liu-Tesche (BLT) (Baum et al.~\cite{baum78}) equation are used to conduct qualitative and quantitative analysis of electromagnetic interactions respectively. Stated in matrix form, BLT equation can be easily programmed and promoted to network BLT equation (Parmantier et al.~\cite{parm90}). After that many authors worked on various problems raised by application of the method. The fundamental BLT equation has been used by Parmantier~(\cite{parm04}) to carry out the topological analysis of a system, and a strategy enlarging the scope of the entire simulation by combining several specific numerical tools has been proposed. In Kirawanich et al.~(\cite{kira05}), a methodology has been proposed to simulate external interaction problems on very large complex electronic systems, and the response of the cables to lightning and electromagnetic pulses has been studied using an EMT-based code, which relies on transmission line theory to determine the transfer function. After that experiment and EMT-based simulation have been performed to study the effect of wideband electromagnetic pulse on a cable behind a slot aperture, and validated the numerical results by comparing with the experimental results (Kirawanich et al.~\cite{kira08}). However such a sequential method cannot solve multiple responses simultaneously. Therefore, it is of great interest to think of developing a hybrid method to quickly predict the electromagnetic interference.

In this paper, we present a hybrid method that combines BLT equation and transfer function to predict the interference response of a radio telescope system. In Section \ref{sect:Ana}, a coupling intensity criterion is proposed by analyzing the conditions which BLT equation simplifies to transfer function. In Section \ref{sect:App}, based on the EMT theory, the coupling paths and volumes of radio telescope system are determined. By the criterion proposed in Section \ref{sect:Ana}, the coupling path of radio telescope is divided into strong coupling and weak coupling sub-paths, and the solving process of system response is given. In Section \ref{sect:The}, the responses of a typical model obtained by the proposed method are compared with those obtained by numerical method. Finally, some conclusions are made in Section \ref{sect:Conclusion}.


\section{Analytical method}
\label{sect:Ana}

\subsection{Generalized BLT equation}
\label{sect:Gen}

The conventional BLT equation is a frequency domain method for solving the interference response of transmission-line load (Tesche \& Liu~\cite{tesc86}). Similar to the propagation of electromagnetic field along the transmission line, the electromagnetic field in free space can also be described with incident wave and reflected wave, so it can be regarded as a virtual transmission line. Therefore, the generalized BLT equation can be established to solve the voltage response of transmission line and the electric field response of the space simultaneously.

The flux of interference into the entire system can be described with a topological network. The network is constituted by tubes related to each other through junctions.
A propagation equation can be defined to describe the relationship between reflected wave $ W_{i,m} ^{ref} $ and incident wave $ W_{i,n} ^{inc} $ at each extremity of tube $ i $:
\begin{equation}
W_{i,m} ^{ref}=P_{m:n}^{i} \cdot W_{i,n} ^{inc}-W_{s_{i}}
\end{equation}

The scattering equation is thereby defined to describe the relationship between each reflected wave $ W_{i,m} ^{ref} $ and incident wave $ W_{j,m} ^{inc} $ at junction $ J_{m} $:
\begin{equation}
W_{i,m} ^{ref}=S_{i:j}^{m} \cdot W_{j,m} ^{inc}
\end{equation}
Where $ P_{m:n}^{i} $ and $ S_{i:j}^{m} $ are propagation and scattering parameters, respectively. $ W_{s_{i}} $ is excitation source applied on tube $ i $.
By establishing propagation and scattering supermatrices involving all the waves on the network, the propagation and scattering equations of the entire network can be defined as:
\begin{equation}
\emph{\textbf{W}}^{ref}=\emph{\textbf{P}} \cdot \emph{\textbf{W}}^{inc}-\emph{\textbf{W}}_{s}
\end{equation}
\begin{equation}
\emph{\textbf{W}}^{ref}=\emph{\textbf{S}} \cdot \emph{\textbf{W}}^{inc}
\end{equation}
Where $ \emph{\textbf{W}}^{inc} $, $ \emph{\textbf{W}}^{ref} $ and $ \emph{\textbf{W}}_{s} $ are incident wave, reflected wave and excitation source supervectors, respectively, $ \emph{\textbf{P}} $ and $ \emph{\textbf{S}} $ are the propagation and scattering supermatrices of the network, respectively. The generalized BLT equation of the entire network is:
\begin{equation}
\emph{\textbf{W}}=\emph{\textbf{W}}^{inc}+\emph{\textbf{W}}^{ref}=(\emph{\textbf{E}}+\emph{\textbf{S}})(\emph{\textbf{P}}-\emph{\textbf{S}})^{-1}\emph{\textbf{W}}_{s}
\end{equation}
Where $ \emph{\textbf{E}} $ is the unit supermatrix. By substituting the corresponding parameters into generalized BLT equation, the response of each extremity of the tube can be expressed as $ W_{i,m}=W_{i,m}^{inc}+W_{i,m}^{ref} $, i.e. the voltage or electric field that interference induces on junction $ J_{m} $ through tube $ i $, and the response of each junction can be expressed as $ W_{Jm}=\sum\limits_{i}W_{i,m} $, i.e. the sum of the voltages or electric fields that interference induces on junction $ J_{m} $ through all the tubes connected to this junction.

\subsection{Strong coupling and weak coupling paths}
\label{sect:Str}

When a topological network is solved with generalized BLT equation, the order of propagation and scattering matrices is twice the number of all tubes. Hence, it is very hard to solve the system response with the increase of the coupling paths. Besides, a coupling path of EMI consists of multiple tubes with different coupling modes, so that it is also difficult to solve the response accurately by using a single method. If a coupling path could be divided into sub-paths which can be analyzed respectively, the overall response of the entire system could be obtained by integrating the responses of each sub-path, which would greatly reduce the order of propagation and scattering matrices as well as improve the efficiency of solving the system response.

According to the transmission line theory, the electromagnetic wave in a tube is the superposition of the incident wave and reflected wave, which embodies as bi-directional coupling. Under this condition, by solving the generalized BLT equation, the response of junction $ J_{m} $ can be expressed as:
\begin{equation}
W_{Jm}=\sum_{i}W_{i,m}=\frac{\sum\limits_{j}\left[\left(1+\sum\limits_{i}S_{i:j}^{m}\right)\sum\limits_{k}W_{s_{k}}A_{k:j}\right]}{|\emph{\textbf{P}}-\emph{\textbf{S}}|}
\end{equation}
Where $ S_{i:j}^{m} $ is the scattering parameter of $ J_{m} $, $ k $ is the tube on which excitation source $ W_{s_{k}} $ applied, and $ A_{k:j} $ is the algebraic complement of the element in row and column corresponding to tubes $ k $ and $ j $ of $ |\emph{\textbf{P}}-\emph{\textbf{S}}| $. Because the value of denominator $ |\emph{\textbf{P}}-\emph{\textbf{S}}| $ is associated with the propagation and scattering parameters of all junctions, the response of junction $ J_{m} $ is related to all junctions of the network. In this case, the path between the interference source and the junction to be solved is defined as the strong coupling path.

The response of $ J_{m} $ can also be expressed by:
\begin{equation}
W_{Jm}=\frac{W_{Jm}}{W_{Jz}} \cdot \cdots \cdot \frac{W_{Jy}}{W_{Jx}} \cdot W_{Jx}
\end{equation}

Where the ratio of $ W_{Jy} $ to $ W_{Jx} $ can be expressed as:
\begin{equation}
\frac{W_{Jy}}{W_{Jx}}=\frac{\sum\limits_{i}W_{i,y}}{\sum\limits_{i}W_{i,x}}=\frac{\sum\limits_{j}\left[\left(1+\sum\limits_{i}S_{i:j}^{y}\right)\sum\limits_{k}W_{s_{k}}A_{k:j}\right]}{\sum\limits_{j}\left[\left(1+\sum\limits_{i}S_{i:j}^{x}\right)\sum\limits_{k}W_{s_{k}}A_{k:j}\right]}
\end{equation}

Through the derivation, if the interference is directed from $ J_{x} $ to $ J_{y} $ and the scattering parameter $ S_{i:j}^{y} (i>j) $ or propagation parameter $ P_{y:y+1} $ or $ P_{y+1:y} $ is much less than 1, the electromagnetic wave in $ J_{y} $ has only one direction, which embodies as unidirectional coupling. If the interference is directed from $ J_{y} $ to $ J_{x} $ and the scattering parameter $ S_{i:j}^{x} (i<j) $ or propagation parameter $ P_{x-1:x} $ or $ P_{x:x-1} $ is much less than 1, the same characteristic shows in $ J_{x} $, which also embodies as unidirectional coupling. In these cases, algebraic complement $ A_{k:j} $ can be expressed as the product of three determinants which correspond to the sub-path from interference source to $ J_{x} $, the sub-path from $ J_{x} $ to $ J_{y} $, and the sub-path from $ J_{y} $ to the end of the sub-path. By dividing out the same part, the ratio of $ W_{Jy} $ to $ W_{Jx} $ is only related to the two junctions and the junctions between them. In this case, the path between the interference source and the junction to be solved is defined as the weak coupling path. Described above set a criterion defining a weak coupling path, which is named as coupling intensity criterion. It is to be noted that, if there is a function to describe the bi-directional coupling characteristic of a strong coupling path, the strong coupling path can be handled as a weak coupling path.

From the above analysis, the coupling path can be expressed as a cascade of multiple weak coupling sub-paths and strong coupling sub-paths. By solving the different sub-paths with corresponding methods, the response of the system could be solved more effectively.

\subsection{Transfer function of weak coupling path}
\label{sect:Weak}

Transfer function is the ratio function of response phasor to excitation phasor at different extremities. Transfer function only considers the incidence and propagation of the electromagnetic wave, i.e. unidirectional coupling. Because each weak coupling sub-path is independent, that is, it would not be affected by other sub-paths, a transfer function can be established to describe the relationship between two ends of a weak coupling sub-path. By choosing different outputs and inputs, four transfer functions can be defined to represent the coupling efficiency of the electromagnetic fields between two volumes, the coupling efficiency of the voltages/currents between two points on wires or circuits, and the coupling efficiencies between the electromagnetic field and the voltage/current, respectively. (Qiu et al.~\cite{qiu17}).

The transfer function of the electromagnetic fields between two volumes is established as:
\begin{equation}
(E_{Jy},H_{Jy})=T_{x:y}^{r} \cdot (E_{Jx},H_{Jx})
\end{equation}
Where $ (E_{Jx},H_{Jx}) $ is the incident field in volume $ V_{Jx} $, $ (E_{Jy},H_{Jy}) $ is the coupled field in volume $ V_{Jy} $, $ T_{x:y}^{r} $ is the transfer function of field-to-field coupling from $ V_{Jx} $ to $ V_{Jy} $ (superscript denotes the coupling mode). In the same manner, the transfer function of field-to-wire coupling from volume $ V_{Jx} $ to wire or circuit $ L_{Jy} $, the transfer function of radiation coupling from wire or circuit $ L_{Jx} $ to volume $ V_{Jy} $, and the transfer function of conduction coupling from point $ L_{Jx} $ on wire or circuit to another point $ L_{Jy} $ are established as:
\begin{equation}
(U_{Jy},I_{Jy})=T_{x:y}^{r:c} \cdot (E_{Jx},H_{Jx})
\end{equation}
\begin{equation}
(E_{Jy},H_{Jy})=T_{x:y}^{c:r} \cdot (U_{Jx},I_{Jx})
\end{equation}
\begin{equation}
(U_{Jy},I_{Jy})=T_{x:y}^{c} \cdot (U_{Jx},I_{Jx})
\end{equation}

When the coupling path between $ J_{x} $ and $ J_{m} $ is the cascade of multiple weak coupling sub-paths, the response of junction $ J_{m} $ can be represented as:
\begin{equation}
W_{Jm}=T_{z:m} \cdot \cdots \cdot T_{x:y} \cdot W_{Jx}
\end{equation}

For a complex system whose coupling path contains multiple strong and weak sub-paths, the transfer function of each weak coupling sub-path can be established to solve the output response of the sub-path, then the response treated as excitation source of strong coupling sub-paths is substituted into generalized BLT equation. The final response of the system is given by:
\begin{equation}
\emph{\textbf{W}}=(\emph{\textbf{E}}+\emph{\textbf{S}}')[(\emph{\textbf{P}}'-\emph{\textbf{T}})-\emph{\textbf{S}}']^{-1}\emph{\textbf{W}}_{s}
\end{equation}
Where $ \emph{\textbf{P}}' $ and $ \emph{\textbf{S}}' $ are propagation and scattering supermatrices of strong coupling sub-paths, respectively. $ \emph{\textbf{T}} $ is the transfer supermatrix of interferences which apply on strong coupling sub-paths through the weak coupling sub-paths.

In order to solve the response of the system accurately, it is necessary to describe different sub-paths with the corresponding models and determine the types of transfer functions and the corresponding calculation methods. For example, for the interference which penetrates into the shield, equivalent transmission line theory, equivalent circuit method (Tesche et al.~\cite{tesc97}; Yin \& Du~\cite{yin16}) and scattering theory (Tsang et al.~\cite{tsang00}) can be used; for the interference through aperture, dyadic green's function method (Yang \& Volakis~\cite{yang05}), method of moments (Audone \& Balma~\cite{audo89}; Araneo \& Lovat~\cite{aran09}) and finite-difference time-domain method (Georgakopoulos et al.~\cite{geor01}) can be used; for the interference coupled to the antenna and cable, field-to-wire coupling theory (Agrawal et al.~\cite{agra80}) can be used; for the interference between two cables, crosstalk theory (Mohr~\cite{mohr67}) and spectral analysis theory can be used.

\section{Application of the hybrid method to radio telescope system}
\label{sect:App}

\subsection{Coupling characteristics of interference in radio telescope system}
\label{sect:Cou}

\begin{figure}
\centering
  \includegraphics[width=0.95\textwidth, angle=0]{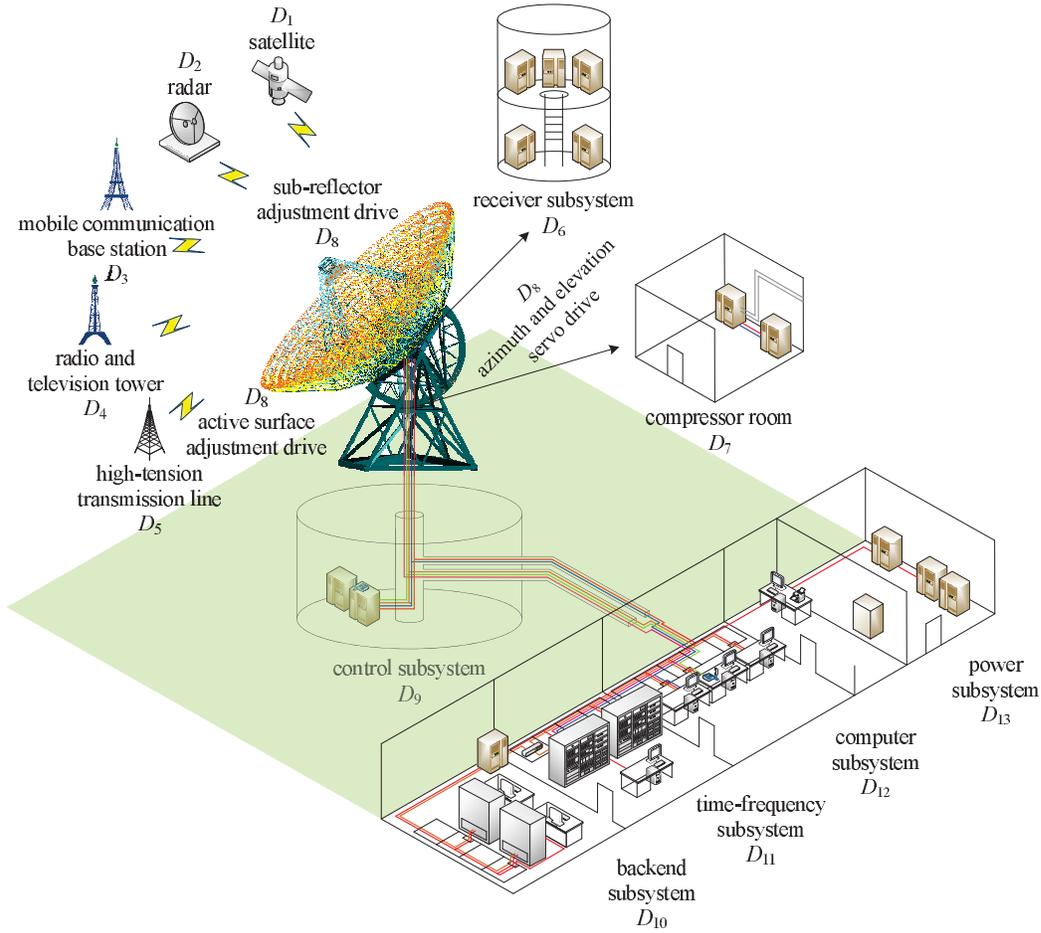}
\caption{Radio telescope system and its electromagnetic environment.}
\label{fig:1}
\end{figure}

Radio telescope system is a complex system which contains high sensitive receivers, high power drive subsystem, control and monitoring subsystem and other subsystems (Wang~\cite{wang14}). The system and its electromagnetic environment are shown in Figure~\ref{fig:1}. Both electromagnetic emission signals of the system in operation and various electromagnetic signals in the environment will bring interference to the radio telescope.

\begin{figure}
\centering
  \includegraphics[width=\textwidth, angle=0]{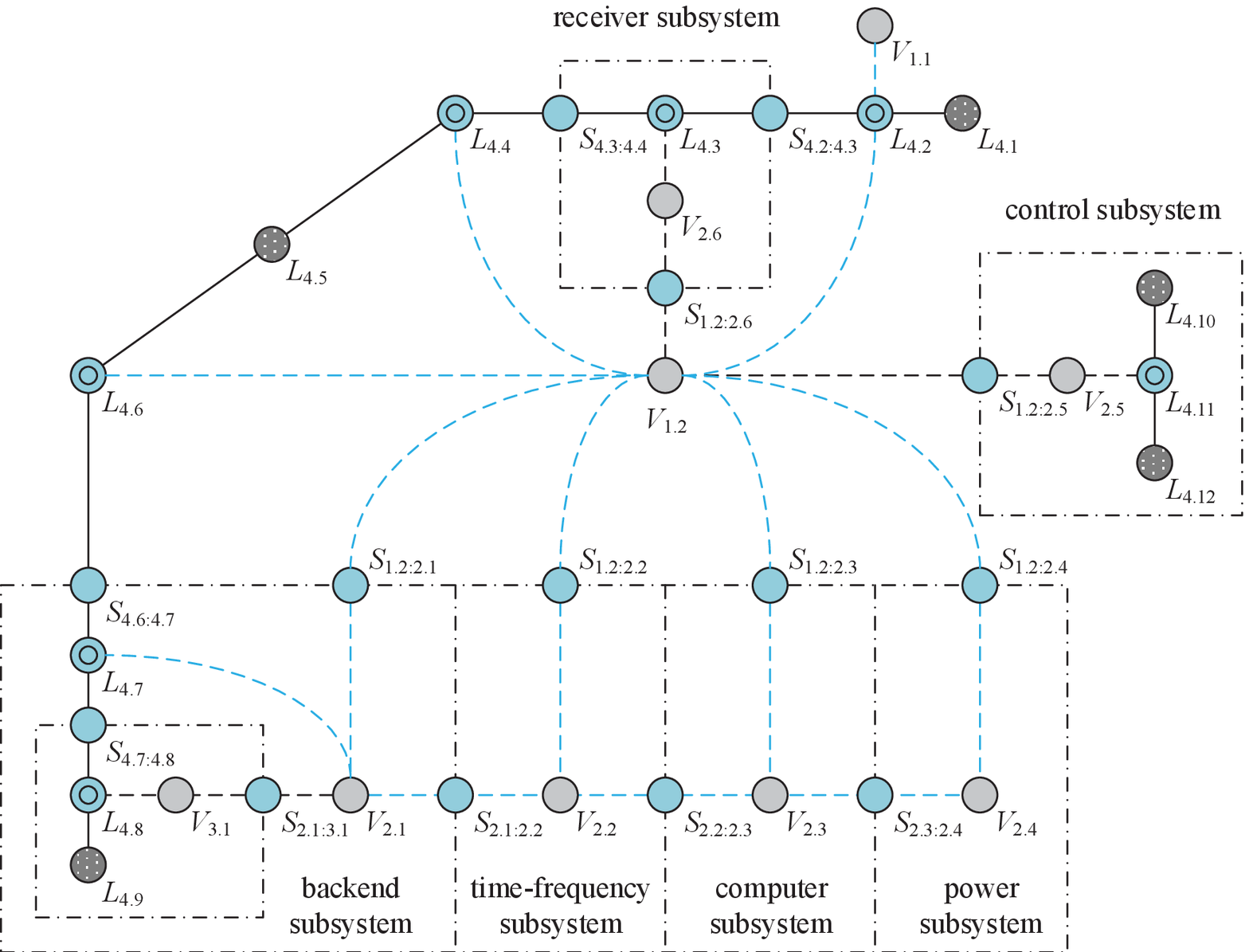}
\caption{Interference sequence diagram of a radio telescope system.}
\label{fig:2}
\end{figure}

The electromagnetic interference to the radio telescope system, essentially, is the interference to sensitive components or circuits of receiver subsystem and control subsystem. Based on the electromagnetic topology theory, the volumes of a radio telescope system are divided. According to the coupling paths of EMI in the system, the interference sequence diagram is obtained, as shown in Figure~\ref{fig:2}. The external interference sources are in volume $ V_{1.1} $, and the internal interference sources are in volume $ V_{2.1} \sim V_{2.5} $. The sensitive circuit of control subsystem is analog circuit, in volume $ V_{2.5} $, and the sensitive component of receiver subsystem is cryogenic amplifier, in volume $ V_{2.6} $. In order to analyze easily, the radio telescope system is simplified, and the interference which penetrates into the shield is ignored.

In Figure~\ref{fig:2}, the coupling path of EMI consists of multiple tubes with different coupling modes, including field-to-antenna coupling, field-to-wire coupling, field-to-field coupling, and conduction coupling. The external interferences and internal interferences can couple into the feed by field-to-antenna coupling, then couple to sensitive components or circuits by conduction coupling. Besides, the internal interferences can also couple into the interior of the receiver subsystem and control subsystem by field-to-field coupling, then couple to the conduction sub-path by field-to-wire coupling.

Because the receiver passively receives electromagnetic radiation of celestial body, considering the principle of dual-reflector antenna, on most occasions, the reflection of the antenna can be ignored. Therefore, the field-to-antenna coupling sub-path can be handled as weak coupling sub-path.

Because of the various structure factors (enclosure, shape of the shield, material, aperture, etc.) and circuit factors (component, wire, grounding method, power, intensity of the circuit, etc.) of radio telescope system, both the field-to-field coupling and the conduction coupling have different coupling strengths. When the interference encounters a metal shield or transfers to a circuit of impedance mismatch, the reflection will occur, and the sub-path can be handled as strong coupling sub-path. On the other hand, when the interference encounters a non-metallic shield or transfers to a circuit of impedance match, the reflection can be ignored, and the sub-path can be handled as weak coupling sub-path. For the field-to-wire coupling, if the length of transmission line is much greater than the distance between the conductors of the transmission line, the secondary radiation of the transmission line can be ignored, and the sub-path can also be handled as weak coupling sub-path.

\subsection{The analysis of subsystem}
\label{sect:Sub}

Through the above analysis, the coupling path of radio telescope system can be divided into strong coupling and weak coupling sub-paths which can be modeled respectively. By establishing transfer functions of interferences, the responses of strong coupling and weak coupling sub-paths can be integrated as the response of the entire system, and the quantitative analysis of interference in radio telescope system is realized.

\begin{figure}
\centering
  \includegraphics[width=\textwidth, angle=0]{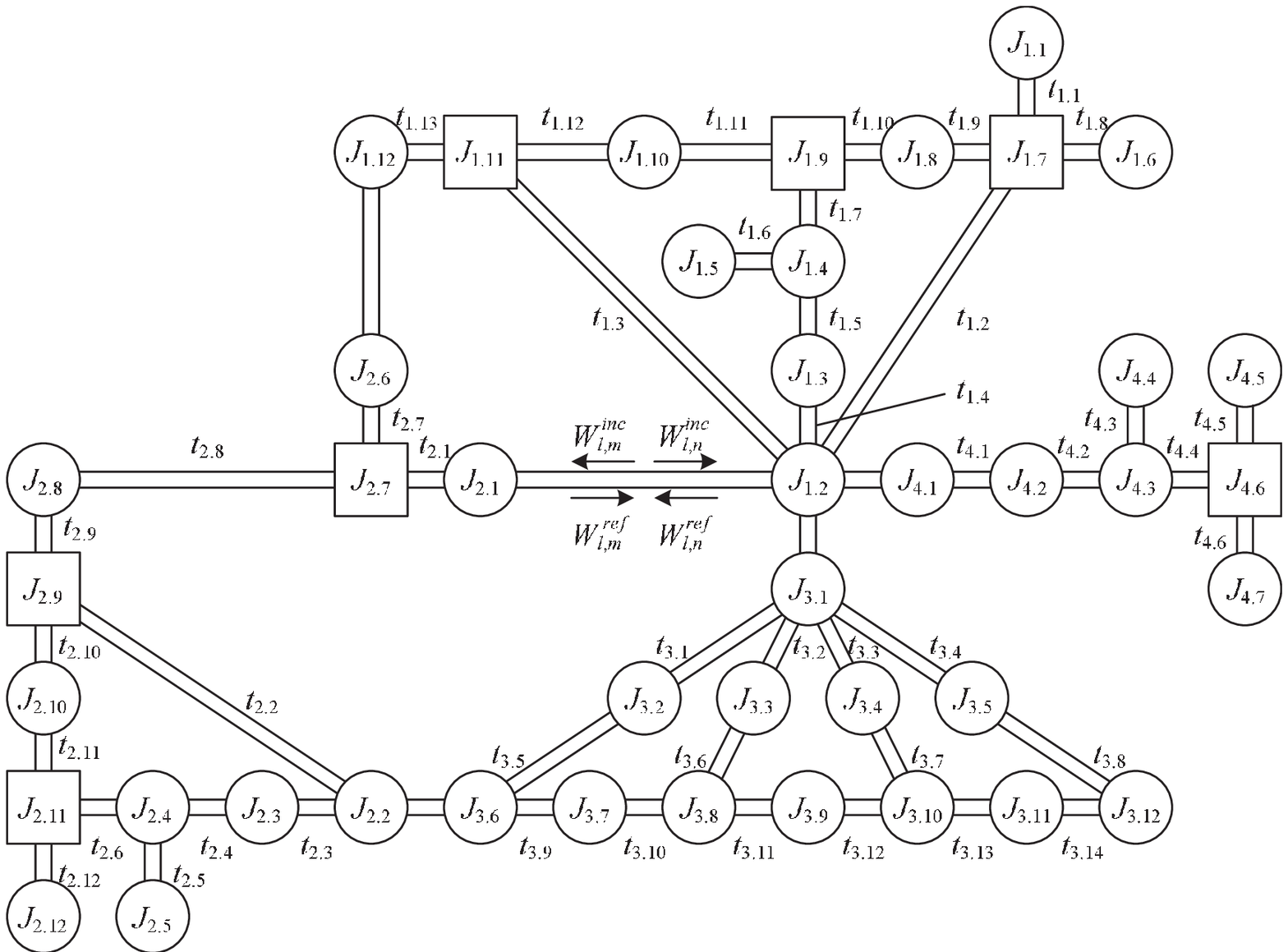}
\caption{Topological network associated to the interaction sequence diagram.}
\label{fig:3}
\end{figure}

According to the dividing criterion of coupling intensity, the coupling path of radio telescope system is divided. Combining with the interference sequence diagram, the topological network of radio telescope system is established, as shown in Figure~\ref{fig:3}. In Figure~\ref{fig:3}, two junctions are connected by a tube which are associated to unidirectional or bi-directional branch.

In Figure~\ref{fig:3}, based on the composition of the system, the radio telescope system is preliminarily divided into four subsystems (the subsystem $ D_{11} \sim D_{13} $ which have the same coupling mode have been combined into one subsystem). The coupling path of each subsystem is composed of strong coupling sub-paths and weak coupling sub-paths, and the coupling paths of two subsystem are connected by weak coupling sub-paths. In addition, junctions $ J_{1.2} $, $ J_{2.1} $, $ J_{3.1} $ and $ J_{4.1} $ represent volume $ V_{1.2} $ in the four subsystems respectively. Similarly, volume $ V_{2.1} $ and wire or circuit $ L_{4.5} $ are represented by two junctions respectively.

In the four subsystems above, the subsystem 1 which completely contains all the typical coupling modes is taken as an example here. In subsystem 1, the sub-paths from external interference region $ V_{1.1} $ ($ J_{1.1} $) and internal interference region $ V_{1.2} $ ($ J_{1.2} $) to feed $ L_{4.2} $ ($ J_{1.7} $), the sub-path from internal interference region $ V_{1.2} $ ($ J_{1.2} $) to wire or circuit $ L_{4.4} $ ($ J_{1.11} $) and the sub-path from the interior of receiver $ V_{2.6} $ ($ J_{1.4} $) to sensitive component $ L_{4.3} $ ($ J_{1.9} $) are classified as weak coupling sub-paths. The sub-path from internal interference region $ V_{1.2} $ ($ J_{1.2} $) through the enclosure of the receiver $ S_{1.2:2.6} $ ($ J_{1.3} $) to the interior of the receiver $ V_{2.6} $ ($ J_{1.4} $) and the sub-path from feed $ L_{4.2} $ ($ J_{1.7} $) through interface $ S_{4.2:4.3} $ ($ J_{1.8} $) to sensitive component $ L_{4.3} $ ($ J_{1.9} $) are classified as strong coupling sub-paths. Among them, $ J_{1.9} $ and $ J_{1.11} $ are the junctions where field-to-wire coupling occurs, and junction $ J_{1.5} $ is the end of the shield.

According to topological network, propagation and scattering supermatrices of strong coupling sub-paths are obtained:
\begin{equation}
\emph{\textbf{P}}=diag(\emph{\textbf{P}}^{1.4},\emph{\textbf{P}}^{1.5},\emph{\textbf{P}}^{1.6},\emph{\textbf{P}}^{1.8},\emph{\textbf{P}}^{1.9},\emph{\textbf{P}}^{1.10},\emph{\textbf{P}}^{1.11},\emph{\textbf{P}}^{1.12},\emph{\textbf{P}}^{1.13})
\end{equation}
\begin{equation}
\emph{\textbf{S}}=diag(S_{1.4:1.4}^{1.2},\emph{\textbf{S}}^{1.3},\emph{\textbf{S}}^{1.4},S_{1.6:1.6}^{1.5},S_{1.8:1.8}^{1.6},\emph{\textbf{S}}^{1.7},\emph{\textbf{S}}^{1.8},\emph{\textbf{S}}^{1.9},\emph{\textbf{S}}^{1.10},\emph{\textbf{S}}^{1.11},S_{1.13:1.13}^{1.12})
\end{equation}
Where $ \emph{\textbf{S}}^{m}=\left[\begin{array}{cc}S_{i:i}^{m}& S_{i:j}^{m}\\S_{j:i}^{m}& S_{j:j}^{m}\end{array}\right] $ is the scattering matrix of junction $ J_{m} $, $ \emph{\textbf{P}}^{i}=\left[\begin{array}{cc}0& P_{m:n}^{i}\\P_{n:m}^{i}& 0\end{array}\right] $ is the propagation matrix between $ J_{m} $ and $ J_{n} $ connected by tube $ i $. Furthermore, transfer supermatrix of interference which applies on conduction sub-path through field-to-wire coupling is obtained:
\begin{equation}
\emph{\textbf{T}}=diag(0,0,0,\emph{\textbf{T}}^{1.4\sim1.10},0,0,0)
\end{equation}
\begin{equation}
\emph{\textbf{T}}^{1.4\sim1.10}=
\left[\begin{array}{lllll;{2pt/2pt}cc}
\multicolumn{5}{c;{2pt/2pt}}{\raisebox{0ex}[0pt]{\Large0}}& & \multicolumn{1}{c}{\raisebox{0ex}[0pt]{\Large0}}\\
\hdashline[2pt/2pt]
(1+S_{1.5:1.5}^{1.4}) \cdot T_{1.4:1.8}^{1.10} & & & (1+S_{1.6:1.6}^{1.4}) \cdot T_{1.4:1.8}^{1.10} & & & \multirow{4}{*}{\raisebox{0ex}[0pt]{\Large0}}\\
(1+S_{1.5:1.5}^{1.4}) \cdot T_{1.4:1.9}^{1.10} & & & (1+S_{1.6:1.6}^{1.4}) \cdot T_{1.4:1.9}^{1.10} & & &\\
(1+S_{1.5:1.5}^{1.4}) \cdot T_{1.4:1.9}^{1.11} & & & (1+S_{1.6:1.6}^{1.4}) \cdot T_{1.4:1.9}^{1.11} & & &\\
(1+S_{1.5:1.5}^{1.4}) \cdot T_{1.4:1.10}^{1.11}& & & (1+S_{1.6:1.6}^{1.4}) \cdot T_{1.4:1.10}^{1.11}& & &
\end{array}\right]
\end{equation}
Where $ T_{m:n}^{i} $ is the transfer function of field-to-wire coupling from junction $ J_{m} $ to the extremity where tube $ i $ connects to junction $ J_{n} $. Substituting the supermatrices into generalized BLT equation, the response of subsystem 1 is obtained:
\begin{equation}
\begin{aligned}
\emph{\textbf{W}}=(&W_{1.4,1.2},W_{1.4,1.3},W_{1.5,1.3},W_{1.5,1.4},W_{1.6,1.4},W_{1.6,1.5},\\
&W_{1.8,1.6},W_{1.8,1.7},W_{1.9,1.7},W_{1.9,1.8},W_{1.10,1.8},W_{1.10,1.9},\\
&W_{1.11,1.9},W_{1.11,1.10},W_{1.12,1.10},W_{1.12,1.11},W_{1.13,1.11},W_{1.13,1.12})^{T}
\end{aligned}
\end{equation}

The response of junction $ J_{1.9} $ can be represented as:
\begin{equation}
\begin{aligned}
W_{J1.9}&=W_{1.10,1.9}\!+\!W_{1.11,1.9}\\
&=\frac{(1\!+\!S_{1.10:1.10}^{1.9}\!+\!S_{1.11:1.10}^{1.9})\sum\limits_{k}W_{s_{k}}A_{k:1.10}\!+\!(1\!+\!S_{1.10:1.11}^{1.9}\!+\!S_{1.11:1.11}^{1.9})\sum\limits_{k}W_{s_{k}}A_{k:1.11}}{|\emph{\textbf{P}}-\emph{\textbf{S}}|}
\end{aligned}
\end{equation}

Considering only the coupling path from junction $ J_{1.2} $ to sensitive component through aperture, the response of junction $ J_{1.9} $ can be represented as:
\begin{equation}
W_{J1.9}=\frac{(1\!+\!S_{1.10:1.10}^{1.9}\!+\!S_{1.11:1.10}^{1.9})W_{1.4}A_{1.4:1.10}\!+\!(1\!+\!S_{1.10:1.11}^{1.9}\!+\!S_{1.11:1.11}^{1.9})W_{1.4}A_{1.4:1.11}}{|\emph{\textbf{P}}-\emph{\textbf{S}}|}
\end{equation}

By means of simulation and measurement, the propagation and scattering parameters of each junction in the subsystem can be obtained, then the response of the subsystem can be solved. When the junction to be solved and the interference source are in the same subsystem, only this subsystem needs to be solved. But when they are in different subsystems, the transfer function between subsystems needs to be built to solve the response of the merged path synthetically.

\subsection{The analysis of system}
\label{sect:Sys}

Because multiple interference sources couple to sensitive component of receiver subsystem by multiple paths and each path can be expressed as the cascade of strong coupling and weak coupling sub-paths, according to the superposition principle of the circuit, the final response of junction $ J_{1.9} $ can be represented as:
\begin{equation}
\begin{aligned}
&(U_{J1.9},I_{J1.9})=T_{1.1:1.7}^{r:c}\!\cdot\!M_{1.7:1.9}^{c}\!\cdot\!(E_{J1.1},H_{J1.1})\\
&+\!M_{3.6:3.1}^{r}\!\cdot\!T_{3.1:1.2}^{r}\!\cdot\!(T_{1.2:1.7}^{r:c}\!\cdot\!M_{1.7:1.9}^{c}\!+\!M_{1.2:1.4}^{r}\!\cdot\!T_{1.4:1.9}^{r:c}\!+\! T_{1.2:1.11}^{r:c}\!\cdot\!M_{1.11:1.9}^{c})\!\cdot\!(E_{J3.6},H_{J3.6})\\
&+\!M_{3.8:3.1}^{r}\!\cdot\!T_{3.1:1.2}^{r}\!\cdot\!(T_{1.2:1.7}^{r:c}\!\cdot\!M_{1.7:1.9}^{c}\!+\!M_{1.2:1.4}^{r}\!\cdot\!T_{1.4:1.9}^{r:c}\!+\! T_{1.2:1.11}^{r:c}\!\cdot\!M_{1.11:1.9}^{c})\!\cdot\!(E_{J3.8},H_{J3.8})\\
&+\!M_{3.10:3.1}^{r}\!\cdot\!T_{3.1:1.2}^{r}\!\cdot\!(T_{1.2:1.7}^{r:c}\!\cdot\!M_{1.7:1.9}^{c}\!+\!M_{1.2:1.4}^{r}\!\cdot\!T_{1.4:1.9}^{r:c}\!+\! T_{1.2:1.11}^{r:c}\!\cdot\!M_{1.11:1.9}^{c})\!\cdot\!(E_{J3.10},H_{J3.10})\\
&+\!M_{3.12:3.1}^{r}\!\cdot\!T_{3.1:1.2}^{r}\!\cdot\!(T_{1.2:1.7}^{r:c}\!\cdot\!M_{1.7:1.9}^{c}\!+\!M_{1.2:1.4}^{r}\!\cdot\!T_{1.4:1.9}^{r:c}\!+\! T_{1.2:1.11}^{r:c}\!\cdot\!M_{1.11:1.9}^{c})\!\cdot\!(E_{J3.12},H_{J3.12})\\
&+\!M_{4.3:4.1}^{r}\!\cdot\!T_{4.1:1.2}^{r}\!\cdot\!(T_{1.2:1.7}^{r:c}\!\cdot\!M_{1.7:1.9}^{c}\!+\!M_{1.2:1.4}^{r}\!\cdot\!T_{1.4:1.9}^{r:c}\!+\! T_{1.2:1.11}^{r:c}\!\cdot\!M_{1.11:1.9}^{c})\!\cdot\!(E_{J4.3},H_{J4.3})
\end{aligned}
\end{equation}
Where $ M_{m:n} $ represents the strong coupling sub-path, and $ T_{m:n} $ represents the weak coupling sub-path. According to the above process, by establishing the generalized BLT equation of strong coupling sub-paths and the transfer functions of weak coupling sub-paths and substituting the excitation sources, the final response of junction $ J_{1.9} $ can be obtained. In this case, the order of established propagation and scattering supermatrices is reduced from 90 to 46, where the 90-order supermatrices correspond to all the paths of the topological network in Figure~\ref{fig:3}, and the 46-order matrices correspond to the strong coupling sub-paths between the interference source and the junction to be solved. It is observed that this method can reduce the order of matrices to be solved, thereby reducing the computation time and the difficulty of solving. In the same manner, the interference coupled to control subsystem can also be analyzed.

\section{Numerical results and discussion}
\label{sect:The}
In this section, similar to the coupling path through aperture of subsystem 1 described in section \ref{sect:Sub}, a typical model of a two-conductor transmission line inside a rectangular shielding cavity with aperture is considered. The proposed hybrid method is used to solve the electric field response of an observation point in the cavity and the induced current of transmission-line load, and the accuracy is verified by comparing the responses with the results obtained by {\sc cst}.

\begin{figure}
\centering
  \includegraphics[width=0.6\textwidth, angle=0]{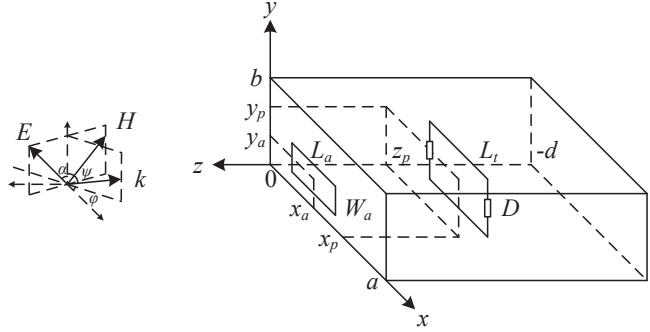}
\caption{Model of transmission line inside rectangular shielding cavity with aperture.}
\label{fig:4}
\end{figure}

The geometry of the model is shown in Figure~\ref{fig:4}. Inside dimensions of the rectangular cavity are 300\,mm$\times$120\,mm$\times$260\,mm, and the thickness is $t=1$\,mm. A rectangular aperture with dimensions of 40\,mm$\times$20\,mm is on the front wall. The length of transmission line is $L_{t}=100$\,mm, the radius of each conductor is $a_{1}=a_{2}=0.5$\,mm, and the distance between two conductors is $D=10$\,mm. The distance between transmission line and aperture is $q=215$\,mm. Point $ P $ of coordinates $ (245,85,-215) $ denotes the observation point located in the center of the transmission line. The incident plane wave propagates along the negative $z$-axis and has an E-filed oriented along the $y$-axis with the amplitude $ V_{0}=1$\,V m$^{-1} $ and with the frequency from 0 to 3\,GHz.

In this model, the coupling path can be expressed as the cascade of field-to-field coupling sub-path, field-to-wire coupling sub-path and conduction coupling sub-path. As described in section \ref{sect:Sub}, the field-to-field coupling sub-path and conduction coupling sub-path are classified as strong coupling sub-paths, and the field-to-wire coupling sub-path is classified as weak coupling sub-path. On the other hand, there is a function that describes the bi-directional coupling characteristic of the field-to-field coupling sub-path, so this sub-path can also be handled as weak coupling sub-path. Therefore, we considered two different scenarios.

\begin{figure*}
\subfloat[Y-axis of the E-field at observation point]{
\begin{minipage}[t]{0.49\textwidth}
\centering
  \includegraphics[width=\textwidth, angle=0]{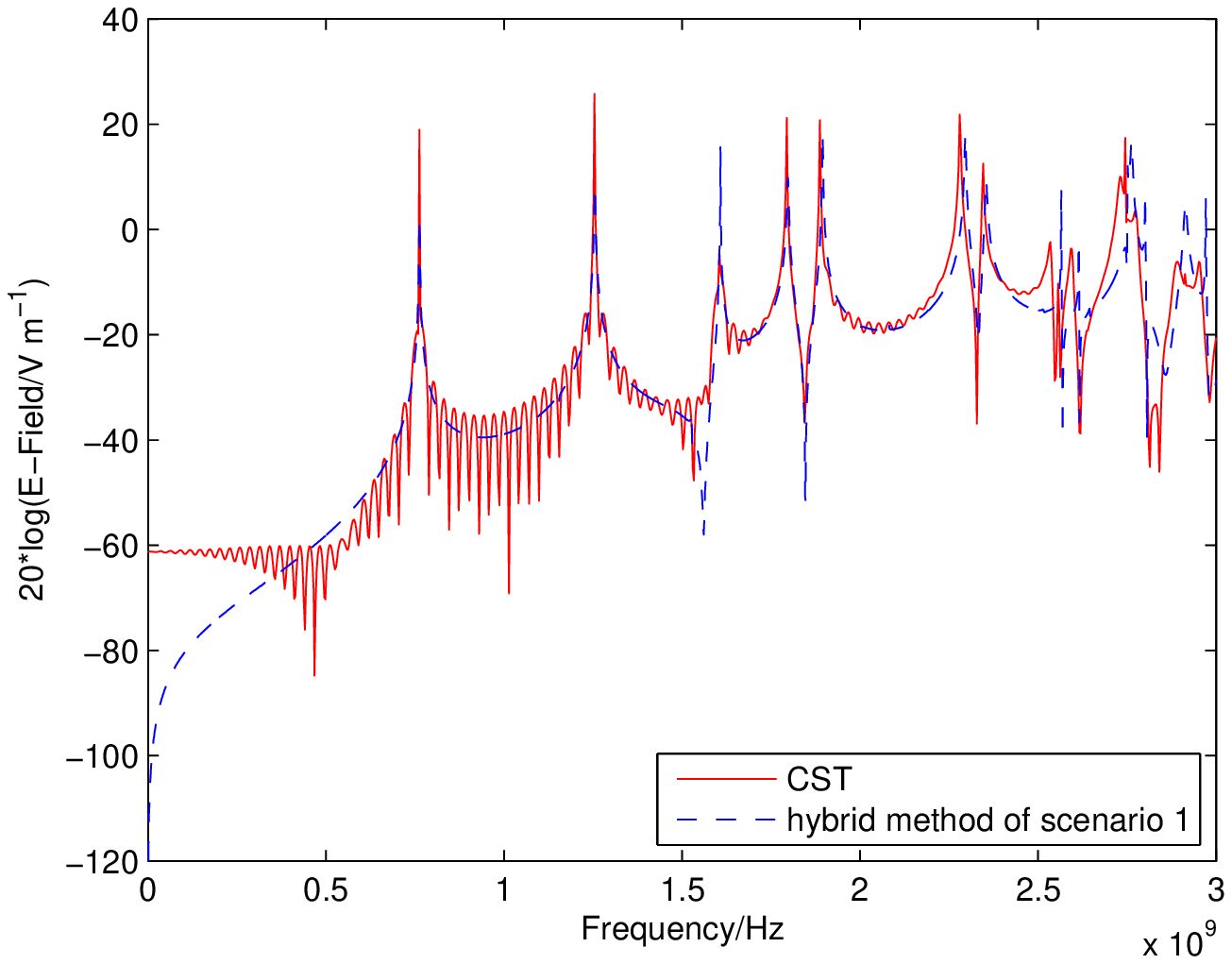}
\end{minipage}}
\subfloat[Induced current of load]{
\begin{minipage}[t]{0.49\textwidth}
\centering
  \includegraphics[width=\textwidth, angle=0]{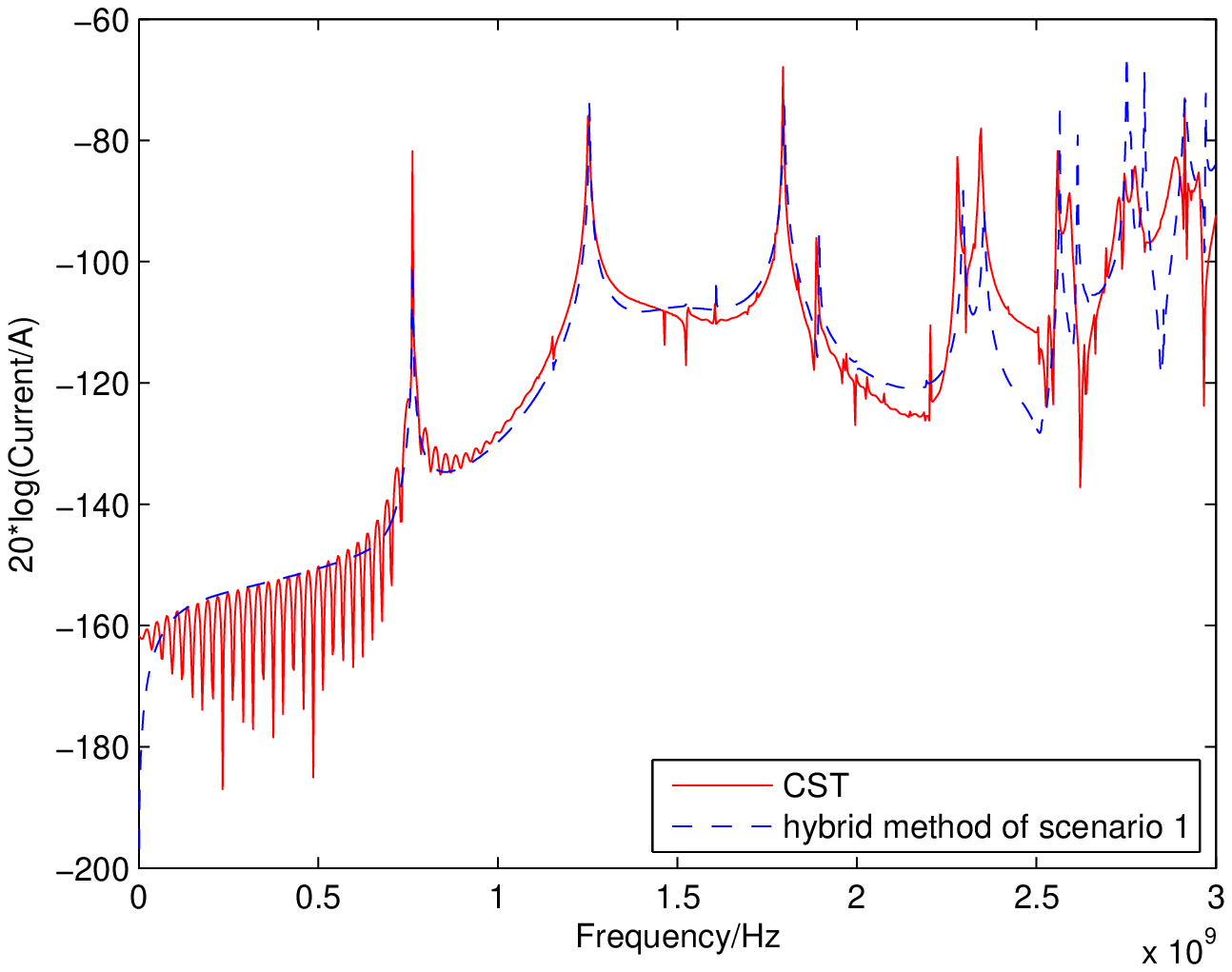}
\end{minipage}}
\caption{Responses solved by {\sc cst} simulation and hybrid method combining BLT equation and transfer function of field-to-wire coupling (scenario 1)}
\label{fig:5}
\end{figure*}

In scenario 1, by establishing the equivalent circuit model, the propagation and scattering parameters of each junction can be obtained to build the BLT equation of the field-to-field coupling and conduction coupling sub-paths. A transfer function of field-to-wire coupling is established and substituted into the BLT equation. In this case, according to Robinson algorithm (Robinson et al.~\cite{robi98}), free space can be represented by a transmission line whose characteristic impedance and propagation constant are $ Z_{0} $ and $ k_{0} $. Aperture can be represented by a coplanar transmission line shorted at the end whose characteristic impedance is $ Z_{oS} $. Cavity can be represented by a rectangular waveguide shorted at the end whose characteristic impedance and propagation constant are $ Z_{g} $ and $ k_{g} $. The two-conductor transmission line in the cavity is uniform and lossless and its characteristic impedance is $ Z_{t} $. By superimposing multiple resonant modes, the y-component of the electric field at arbitrary point $ (x_{p},y_{p},z_{p}) $ in the cavity can be obtained as $ W_{Total}=\sum\limits_{m,n}W_{J}\sin(m \pi x_{p}/a)\cos(n \pi y_{b}/b) $. Figure~\ref{fig:5} shows the responses of the observation point and the load solved by {\sc cst} simulation and hybrid method combining BLT equation and the transfer function of field-to-wire coupling. By comparing with {\sc cst}, it can be seen that the hybrid method has a high precision.

\begin{figure*}
\subfloat[Y-axis of the E-field at observation point]{
\begin{minipage}[t]{0.49\textwidth}
\centering
  \includegraphics[width=\textwidth, angle=0]{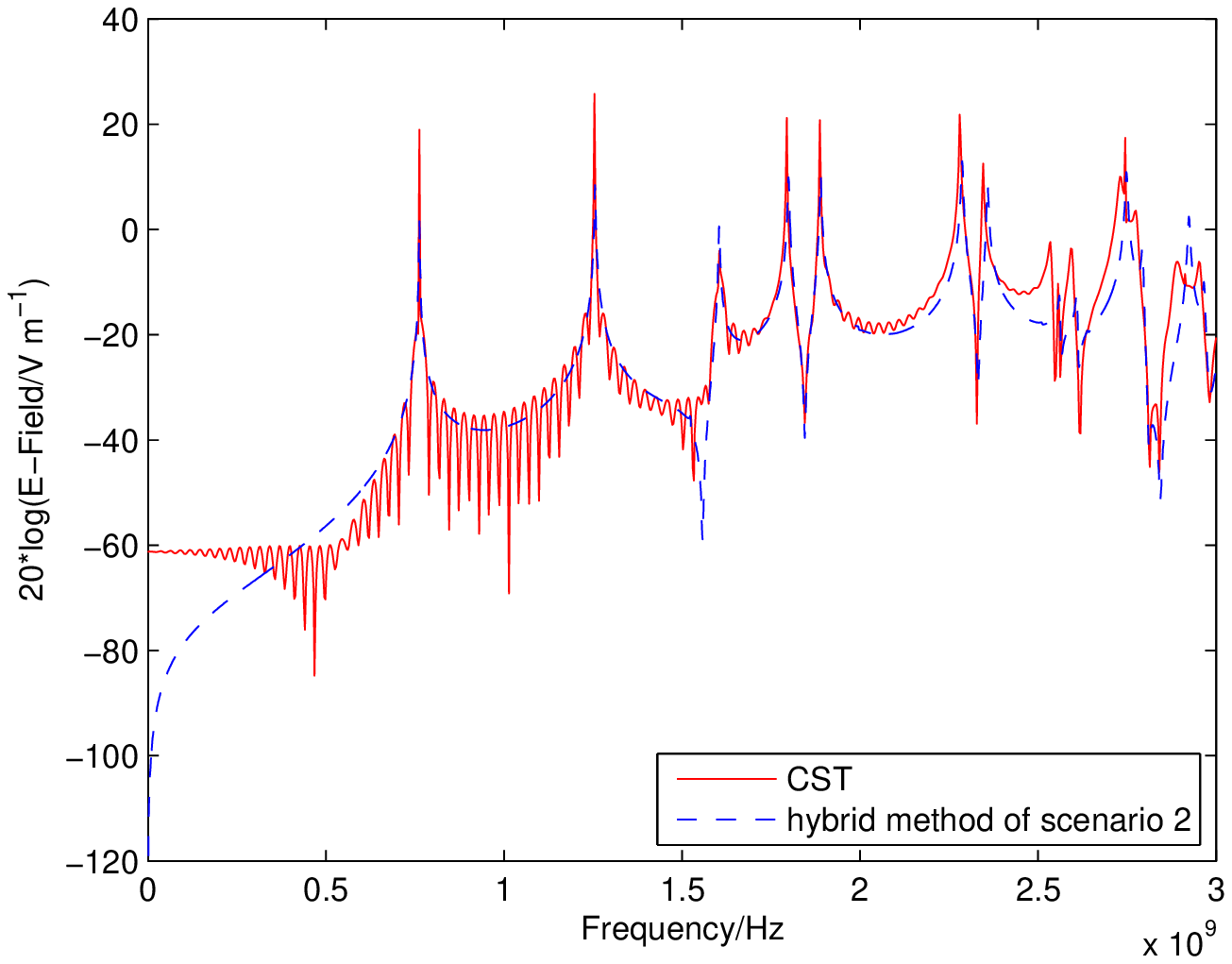}
\end{minipage}}
\subfloat[Induced current of load]{
\begin{minipage}[t]{0.49\textwidth}
\centering
  \includegraphics[width=\textwidth, angle=0]{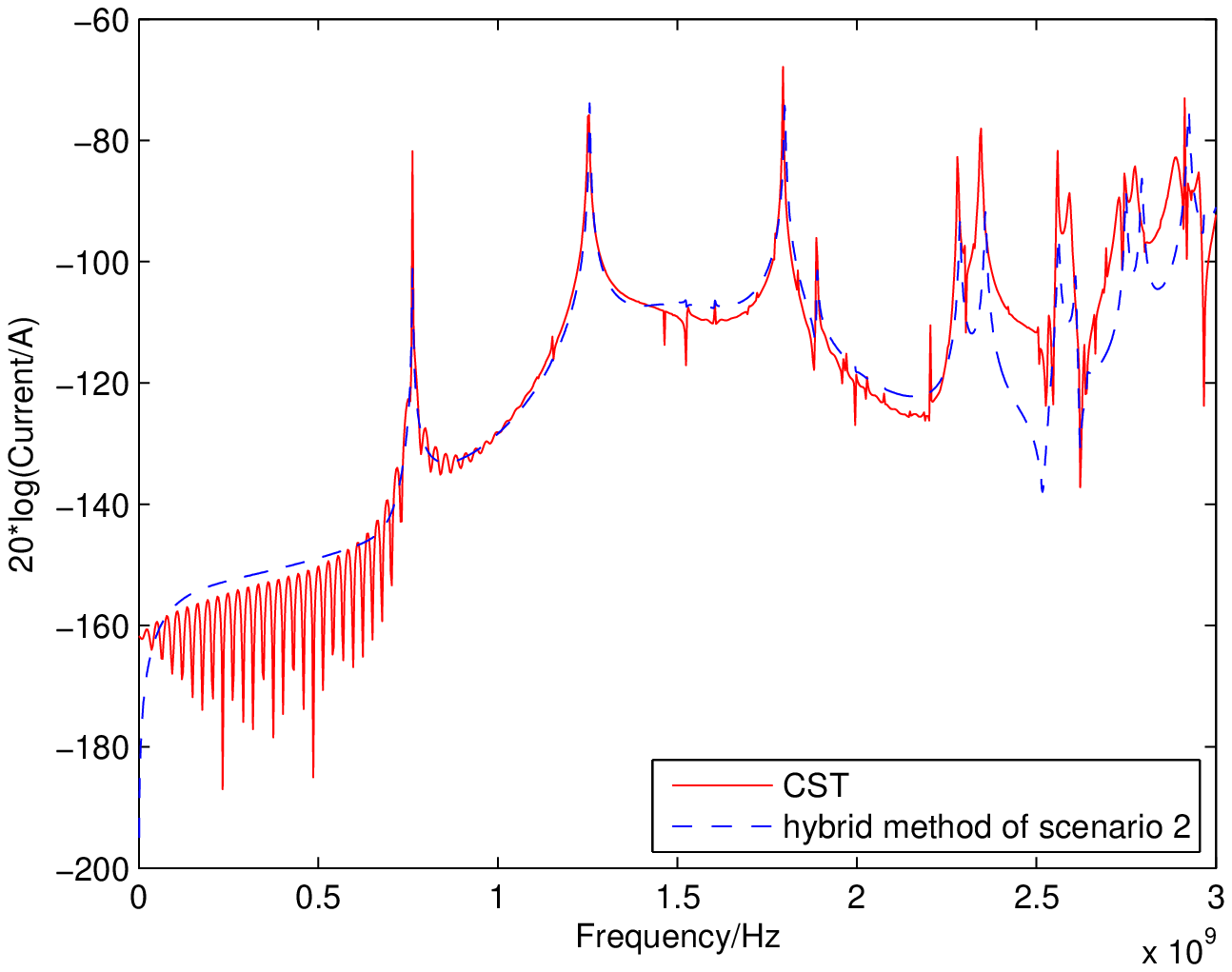}
\end{minipage}}
\caption{Responses solved by {\sc cst} simulation and hybrid method combining BLT equation and transfer function of field-to-field coupling and field-to-wire coupling (scenario 2)}
\label{fig:6}
\end{figure*}

In scenario 2, the BLT equation of field-to-field coupling is replaced by a transfer function. In this case, according to Lee~(\cite{lee86}), the aperture can be represented by equivalent electric dipole and magnetic dipole, then the dyadic Green's function of a rectangular cavity can be used to solve the coupled field inside the cavity. Figure~\ref{fig:6} shows the responses of the observation point and the load solved by {\sc cst} simulation and hybrid method combining BLT equation and transfer function of field-to-field coupling and field-to-wire coupling. It can be seen that, because the equivalent model calculates the radiation field of the magnetic dipole, the results have a higher precision in high frequency.

Table~\ref{Tab:1} shows the time and resource usage taken by {\sc cst} simulation software and the hybrid method programmed with {\sc matlab} to calculate the typical model. All simulations and calculations are performed on the same computer, which has a 3-GHz AMD Athlon(tm) ¢ò X4 640 Processor and 4-GB RAM. It can be seen that the hybrid method coded by {\sc matlab} takes less computing time and resource than {\sc cst}.
\begin{table}
\begin{center}
\caption[]{ Run time and computing resource taken by {\sc cst} simulation and the hybrid method}\label{Tab:1}
 \begin{tabular}{lcc}
\hline\noalign{\smallskip}
Method                      & Run time/s & CPU utility/\%  \\
  \hline\noalign{\smallskip}
{\sc cst}                   & 9712       & 99 \\
hybrid method of scenario 1 & 60.34      & 25 \\
hybrid method of scenario 2 & 6.59       & 25 \\
  \noalign{\smallskip}\hline
\end{tabular}
\end{center}
\end{table}

\section{Conclusion}
\label{sect:Conclusion}
In this paper, a hybrid method for predicting the EMI response of a radio telescope system is proposed. Based on the conditions that BLT equation simplifies to transfer function, the coupling intensity criterion is proposed. Further, the coupling path of radio telescope system is divided into strong coupling and weak coupling sub-paths. Transfer function is used to solve the responses of weak coupling sub-paths, which are treated as the excitation sources of strong coupling sub-paths. Through adopting BLT equation to solve the strong coupling sub-paths, the integrated response of a system is obtained. Finally, the proposed method is used to analyze a simple typical problem. The responses of the observation point and the transmission-line load have been obtained and compared with the results obtained by {\sc cst}. The results show that this hybrid methodology is accurate and efficient to solve the response of the system.

\begin{acknowledgements}
This work was funded by the National Basic Research Program of China under No.2015CB857100 and by the National Natural Science Foundation of China under No.11473061 and No.11103056.
\end{acknowledgements}


\label{lastpage}

\end{document}